\documentclass{article}
\usepackage[keeplastbox]{flushend}
\usepackage{spconf,amsmath,graphicx}
\DeclareMathOperator{\Tr}{Tr}
\usepackage{amssymb}
\usepackage{widetext}
\usepackage{enumitem}
\usepackage{balance}
\usepackage{algorithm}
\usepackage{algorithmic}
\usepackage{color}
\usepackage{tabularx}
\usepackage{multirow}
\usepackage{booktabs}
\usepackage{tabularx}
\usepackage[moderate,lists=normal,indent=normal]{savetrees}
\usepackage{mathtools}
\newcommand{\svdots}{\raisebox{3pt}{$\scalebox{.5}{\vdots}$}} % <- Works
\newcommand{\sddots}{\raisebox{3pt}{$\scalebox{.5}{$\ddots$}$}}

\title{Parametric channel estimation for massive MIMO}
\name{Luc Le Magoarou, St\'ephane Paquelet \thanks{This work has been performed in the framework of the Horizon 2020 project ONE5G (ICT-760809) receiving funds from the European Union. The authors would like to acknowledge the contributions of their colleagues in the project, although the views expressed in this contribution are those of the authors and do not necessarily represent the project.}}
\address{b\raisebox{0.2mm}{\scalebox{0.7}{\textbf{$<>$}}}com, Rennes, France} 
\begin{document}
\ninept
\setlength{\textfloatsep}{1\baselineskip}
\maketitle

\makeatletter
\let\origsection\section
\renewcommand\section{\@ifstar{\starsection}{\nostarsection}}

\newcommand\nostarsection[1]
{\sectionprelude\origsection{#1}\sectionpostlude}
\newcommand\starsection[1]
{\sectionprelude\origsection*{#1}\sectionpostlude}

\newcommand\sectionprelude{%
  \vspace{-0.7em}
}

\newcommand\sectionpostlude{%
  \vspace{-1em}
}
\makeatother

\makeatletter
\let\origsubsection\subsection
\renewcommand\subsection{\@ifstar{\starsubsection}{\nostarsubsection}}

\newcommand\nostarsubsection[1]
{\subsectionprelude\origsubsection{#1}\subsectionpostlude}

\newcommand\starsubsection[1]
{\subsectionprelude\origsubsection*{#1}\subsectionpostlude}

\newcommand\subsectionprelude{%
  \vspace{-0.5em}
}

\newcommand\subsectionpostlude{%
  \vspace{-0.7em}
}
\makeatother
\begin{abstract}
Channel state information is crucial to achieving the capacity of multi-antenna (MIMO) wireless communication systems. It requires estimating the channel matrix. This estimation task is studied, considering a sparse physical channel model, as well as a general measurement model taking into account hybrid architectures. The contribution is twofold. First, the  Cram\'er-Rao bound in this context is derived. Second, interpretation of the Fisher Information Matrix structure allows to assess the role of system parameters, as well as to propose asymptotically optimal and computationally efficient estimation algorithms.
\end{abstract}


\section{Introduction}
\label{sec:intro}
Multiple-Input Multiple-Output (MIMO) wireless communication systems allow for a dramatic increase in channel capacity, by adding the spatial dimension to the classical time and frequency ones \cite{Telatar1999,Tse2005}. This is done by sampling space with several antenna elements, forming antenna arrays both at the transmitter (with $n_t$ antennas) and receiver (with $n_r$ antennas). Capacity gains over single antenna systems are at most proportional to $\min(n_r,n_t)$.

Millimeter wavelengths have recently appeared as a viable solution for the fifth generation (5G) wireless communication systems \cite{Rappaport2013,Swindlehurst2014}. Indeed, smaller wavelengths allow to densify half-wavelength separated antennas, resulting in higher angular resolution and capacity for a given array size. This observation has given rise to the \emph{massive MIMO} field, i.e.\ the study of systems with up to hundreds or even thousands of antennas.

Massive MIMO systems are very promising in terms of capacity. However, they pose several challenges to the research community \cite{Rusek2013,Larsson2014}, in particular for channel estimation. Indeed, maximal capacity gains are obtained in the case of perfect knowledge of the channel state by both the transmitter and the receiver. The estimation task amounts to determine a complex gain between each transmit/receive antenna pair, the narrowband (single carrier) MIMO channel as a whole being usually represented as a complex matrix $\mathbf{H} \in \mathbb{C}^{n_r \times n_t}$ of such complex gains. Without a parametric model, the number of real parameters to estimate is thus $2n_rn_t$, which is very large for massive MIMO systems. %This makes channel estimation challenging, especially in Frequency Division Duplex (FDD) \cite{Bjornson2016}, where the channel has to be estimated for both downlink and uplink.

\noindent{\bf Contributions and organization.} In this work, massive MIMO channel estimation is studied, and its \emph{performance limits} are sought, as well as their dependency on key system parameters. 
\noindent In order to answer this question, the framework of parametric estimation \cite{Kay1993} is used. A physical channel model is first presented, with the general considered observation model, and the objective is precisely stated.
% and classically decompose the mean squared error into bias and variance terms. 
The Cram\'er-Rao bound for is then derived, which bounds the variance of any unbiased estimator. Then, the interpretation of the bound allows to precisely assess the role of system design on estimation performance, as well as to propose new computationally efficient channel estimation algorithms showing asymptotic performance equivalent to classical ones based on sparse recovery.
% Finally, we discuss ways to bound the bias term and analyze the dependence of the bias on the model sparsity. 
%We end the paper by discussing the consequences and possible outcomes of our study.

\section{Problem formulation}
\label{sec:pb}

\noindent{\bf Notations.} Matrices and vectors are denoted by bold upper-case and lower-case letters: $\mathbf{A}$ and $\mathbf{a}$ (except 3D ``\emph{spatial}'' vectors that are denoted $\overrightarrow{a}$); the $i$th column of a matrix $\mathbf{A}$ by: $\mathbf{a}_i$; its entry at the $i$th line and $j$th column by: $a_{ij}$ or $\mathbf{A}_{ij}$. %The $i$th row of a matrix $\mathbf{A}$ is denoted by: $\mathbf{a}^i$ (it is a column vector).
A matrix transpose, conjugate and transconjugate is denoted by: $\mathbf{A}^T$, $\mathbf{A}^*$ and $\mathbf{A}^H$ respectively. The image, rank and trace of a linear transformation represented by $\mathbf{A}$ are denoted: $\text{im}(\mathbf{A})$, $\text{rank}(\mathbf{A})$ and $\text{Tr}(\mathbf{A})$ respectively.
%Sets are denoted by calligraphic symbols: $\mathcal{A}$.
For matrices $\mathbf{A}$ and $\mathbf{B}$, $\mathbf{A}\geq \mathbf{B}$ means that $\mathbf{A}-\mathbf{B}$ is positive semidefinite.
 The linear span of a set of vectors $\mathcal{A}$ is denoted: $\text{span}(\mathcal{A})$. The Kronecker product, standard vectorization and diagonalization operators are denoted by $\text{vec}(\cdot)$, $\text{diag}(\cdot)$, and $\otimes$ respectively. The identity matrix, the $m \times n$ matrix of zeros and ones are denoted by $\mathbf{Id}$, $\mathbf{0}_{m\times n}$ and $\mathbf{1}_{m\times n}$ respectively. $\mathcal{CN}(\boldsymbol\mu,\boldsymbol{\Sigma})$ denotes the standard complex gaussian distribution with mean $\boldsymbol\mu$ and covariance $\boldsymbol{\Sigma}$. $\mathbb{E}(.)$ denotes expectation and $\text{cov}(.)$ the covariance of its argument. 

\subsection{Parametric physical channel model}
\label{ssec:model}
Consider a narrowband block fading channel between a transmitter and a receiver with respectively $n_t$ and $n_r$ antennas. It is represented by the matrix $\mathbf{H} \in \mathbb{C}^{n_r \times n_t}$, in which $h_{ij}$ corresponds to the channel between the $j$th transmit and $i$th receive antennas.

Classically, for MIMO systems with few antennas, i.e.\ when the quantity $n_rn_t$ is small (up to a few dozens), estimators such as the Least Squares (LS) or the Linear Minimum Mean Squared Error (LMMSE) are used \cite{Biguesh2006}.

However, for massive MIMO systems, the quantity $2n_rn_t$ is large (typically several hundreds), and resorting to classical estimators may become computationally intractable. In that case, a parametric model may be used. Establishing it consists in defining a set of $n_p$ parameters $\boldsymbol{\theta} \triangleq (\theta_1,\dots,\theta_{n_p})^T$ that describe the channel as $\mathbf{H} \approx f(\boldsymbol{\theta})$ for a given function $f$, where the approximation is inherent to the model structure and neglected in the sequel (considering $\mathbf{H} = f(\boldsymbol{\theta})$). Channel estimation then amounts to estimate the parameters $\boldsymbol{\theta}$ instead of the channel matrix $\mathbf{H}$ directly. The parametrization is particularly useful if $n_p \ll 2n_rn_t$, without harming accuracy of the channel description. Inspired by the physics of wave propagation under the plane waves assumption, it has been proposed to express the channel matrix as a sum of rank-1 matrices, each corresponding to a single physical path between transmitter and receiver \cite{Sayeed2002}. Adopting this kind of modeling and generalizing it to take into account any three-dimensional antenna array geometry, channel matrices take the form
\begin{equation}
\mathbf{H} = \sum_{p=1}^Pc_p\mathbf{e}_r(\overrightarrow{u_{r,p}}).\mathbf{e}_t(\overrightarrow{u_{t,p}})^H,
\label{eq:channel_model}
\end{equation}
where $P$ is the total number of considered paths (no more than a few dozens), $c_p \triangleq \rho_p \mathrm{e}^{\mathrm{j}\phi_p}$ is the complex gain of the $p$th path, $\overrightarrow{u_{t,p}}$ is the unit vector corresponding to its Direction of Departure (DoD) and $\overrightarrow{u_{r,p}}$ the unit vector corresponding to its Direction of Arrival (DoA). Any unit vector $\overrightarrow{u}$ is described in spherical coordinates by an azimuth angle $\eta$ and an elevation angle $\psi$. The complex response and steering vectors $\mathbf{e}_r(\overrightarrow{u})\in \mathbb{C}^{n_r}$ and $\mathbf{e}_t(\overrightarrow{u})\in \mathbb{C}^{n_t}$ are defined as $(\mathbf{e}_x(\overrightarrow{u}))_i = \frac{1}{\sqrt{n_x}} \mathrm{e}^{-\mathrm{j}\frac{2\pi}{\lambda}\overrightarrow{a_{x,i}}.\overrightarrow{u}}$ for $x\in \{r,t\}$. 
%$\mathbf{e}_r(\overrightarrow{u}) \triangleq \frac{1}{\sqrt{n_r}} (\mathrm{e}^{-\mathrm{j}\frac{2\pi}{\lambda}\overrightarrow{a_{r,1}}.\overrightarrow{u}},\dots,\mathrm{e}^{-\mathrm{j}\frac{2\pi}{\lambda}\overrightarrow{a_{r,n_r}}.\overrightarrow{u}})^T$ and $\mathbf{e}_t(\overrightarrow{u}) \triangleq \frac{1}{\sqrt{n_t}} (\mathrm{e}^{-\mathrm{j}\frac{2\pi}{\lambda}\overrightarrow{a_{t,1}}.\overrightarrow{u}},\dots,\mathrm{e}^{-\mathrm{j}\frac{2\pi}{\lambda}\overrightarrow{a_{t,n_t}}.\overrightarrow{u}})^T$
The set $\{\overrightarrow{a_{x,1}},\dots,\overrightarrow{a_{x,n_x}}\}$ gathers the positions of the antennas with respect to the centroid of the considered array (transmit if $x=t$, receive if $x=r$). In order to lighten notations, the matrix $\mathbf{A}_x \triangleq \frac{2\pi}{\lambda}(\overrightarrow{a_{x,1}},\dots,\overrightarrow{a_{x,n_x}}) \in\mathbb{R}^{3\times n_x}$ is introduced. It simplifies the steering/response vector expression to $\mathbf{e}_x(\overrightarrow{u}) = \frac{1}{\sqrt{n_x}} \mathrm{e}^{-\mathrm{j}\mathbf{A}_x^T\overrightarrow{u}}$, where the exponential function is applied component-wise. In order to further lighten notations, the $p$th \emph{atomic channel} is defined as $\mathbf{H}_p \triangleq c_p\mathbf{e}_r(\overrightarrow{u_{r,p}}).\mathbf{e}_t(\overrightarrow{u_{t,p}})^H$, and its vectorized version $\mathbf{h}_p \triangleq \text{vec}(\mathbf{H}_p) \in \mathbb{C}^{n_rn_t}$. Therefore, defining the vectorized channel $\mathbf{h} \triangleq \text{vec}(\mathbf{H})$, yields $\mathbf{h} = \sum_{p=1}^P \mathbf{h}_p$. Note that the channel description used here is very general, as it handles any three-dimensional antenna array geometry, and not only Uniform Linear Arrays (ULA) or Uniform Planar Arrays (UPA) as is sometimes proposed.

%\begin{equation}
%\mathbf{H} = \sum_{q=1}^Q\sum_{p\in \mathcal{C}_q}c_p\mathbf{e}_r(\overrightarrow{u_{r,p}}).\mathbf{e}_t(\overrightarrow{u_{t,p}})^H
%\label{eq:channel_model_cluster}
%\end{equation}

%In such physical channel models, it is classically assumed that paths are grouped into $Q$ clusters, in which the DoDs and DoAs are close \cite{Saleh1987, Wallace2002, Jensen2004}. This corresponds to a partition of the index set into $Q$ subsets: $\{1,\dots,P\} = \bigcup_{q=1}^Q \mathcal{C}_q$ with $\mathcal{C}_{q_1} \cap \mathcal{C}_{q_2} = \emptyset$ if $q_1 \neq q_2$.

In short, the physical channel model can be seen as a parametric model with $\boldsymbol{\theta} = \{\boldsymbol{\theta}^{(p)}\triangleq(\rho_p,\phi_p,\eta_{r,p},\psi_{r,p},\eta_{t,p},\psi_{t,p}),\;\; p=1{\scriptstyle,\dots,}P\}$. There are thus $6P$ real parameters in this model (the complex gain, DoD and DoA of every path are described with two parameters each). Of course, the model is most useful for estimation in the case where $6P\ll 2n_rn_t$, since the number of parameters is thus greatly reduced.

Note that most classical massive MIMO channel estimation methods assume a similar physical model, but discretize a priori the DoDs and DoAs, so  that the problem fits the framework of sparse recovery \cite{Mallat1993,Tropp2007,Bajwa2010}. The approach used here is different, in the sense that no discretization is assumed for the analysis. %However, as will be expressed in section~\ref{ssec:interpretation}, the two approached can be compared.

\subsection{Observation model}
In order to carry out channel estimation, $n_s$ known pilot symbols are sent through the channel by each transmit antenna. The corresponding training matrix is denoted $\mathbf{X} \in \mathbb{C}^{n_t \times n_s}$. The signal at the receive antennas is thus expressed as $\mathbf{HX} + \mathbf{N}$, where $\mathbf{N}$ is a noise matrix with $\text{vec(N)}\sim \mathcal{CN}(0,\sigma^2\mathbf{Id})$. Due to the high cost and power consumption of millimeter wave Radio Frequency (RF) chains, it has been proposed to have less RF chains than antennas in both the transmitter and receiver \cite{Elayach2014,Alkhateeb2014,Heath2016,Sayeed2016}. Such systems are often referred to as \emph{hybrid architectures}. Mathematically speaking, this translates into specific constraints on the training matrix $\mathbf{X}$ (which has to ``sense'' the channel through analog precoders $\mathbf{v}_i \in \mathbb{C}^{n_t}$, $i=1,\dots,n_{\text{RF}}$, $n_{\text{RF}}$ being the number of RF chains on the transmit side), as well as observing the signal at the receiver through analog combiners. Let us denote $\mathbf{w}_j \in \mathbb{C}^{n_r}$, $j=1,\dots,n_c$ the used analog combiners, the observed data is thus expressed in all generality as
\begin{equation}
\mathbf{Y} = \mathbf{W}^H\mathbf{H}\mathbf{X} + \mathbf{W}^H\mathbf{N},
\label{eq:observation_model}
\end{equation}
where $\mathbf{W} \triangleq (\mathbf{w}_1,\dots,\mathbf{w}_{n_c})$ and the training matrix is constrained to be of the form $\mathbf{X} = \mathbf{VZ}$, where $\mathbf{Z} \in \mathbb{C}^{n_{\text{RF}} \times n_{s}}$ is the digital training matrix. %Channel estimation amounts to retrieve $\mathbf{H}$, knowing $\mathbf{Y}$, $\mathbf{X}$, $\mathbf{W}$ and the noise distribution. In case of a parametric model, as we consider in this paper, retrieving $\mathbf{H}$ is equivalent to retrieving the parameters $\boldsymbol{\theta}$.

\subsection{Objective: bounding the variance of unbiased estimators}
In order to assess the fundamental performance limits of channel estimation, the considered performance measure is the relative Mean Squared Error (rMSE). Denoting indifferently $\mathbf{H}(\boldsymbol{\theta}) \triangleq f(\boldsymbol{\theta})$ or $\mathbf{H}$ the true channel ($\mathbf{h}(\boldsymbol{\theta})$ or $\mathbf{h}$ in vectorized form) and $\mathbf{H}(\hat{\boldsymbol{\theta}})\triangleq f(\hat{\boldsymbol{\theta}})$ or $\hat{\mathbf{H}}$ its estimate ($\mathbf{h}(\hat{\boldsymbol{\theta}})$ or $\hat{\mathbf{h}}$ in vectorized form) in order to lighten notations, rMSE is expressed
\begin{equation}
\renewcommand*{\arraystretch}{1.3}
\arraycolsep=0.6pt
\begin{array}{rl}
\text{rMSE} &= \mathbb{E}\Big(\big\Vert \mathbf{H}- \hat{\mathbf{H}} \big\Vert_F^2\Big).\big\Vert \mathbf{H} \big\Vert_F^{-2} \\
&=\Big(\underbrace{\Tr\Big(\text{cov}\big(\hat{\mathbf{h}}\big)\Big)}_{\text{Variance}} + \underbrace{\big\Vert \mathbb{E}(\hat{\mathbf{H}}) - \mathbf{H}\big\Vert_F^2}_{\text{Bias}}\Big).\big\Vert \mathbf{H} \big\Vert_F^{-2}, \\
\end{array}
%\label{eq:MSE}
\label{eq:MSE_decomposed}
\end{equation}
where the bias/variance decomposition can be done independently of the considered model \cite{Kay1993}.
%\begin{equation}
%\text{MSE} = 
%\underbrace{\Tr\Big(\text{cov}\big(\hat{\mathbf{h}}\big)\Big)}_{\text{Variance}} + \underbrace{\big\Vert \mathbb{E}(\hat{\mathbf{H}}) - \mathbf{H}\big\Vert_F^2}_{\text{Bias}}.
%\label{eq:MSE_decomposed}
%\end{equation}
The goal here is to lower-bound the variance term, considering the physical model introduced in the previous subsection. The bias term is not studied in details here, but its role is evoked in section~\ref{ssec:general_bound}. %This means we make the assumption that the model allows to perfectly describe channel matrices.

\section{Cram\'er-Rao lower bound}
\label{sec:variance}
In this section, the variance term of eq.~\eqref{eq:MSE_decomposed} is bounded using the Cram\'er-Rao Bound (CRB) \cite{Rao1945,Cramer1946}, which is valid for any unbiased estimator $\hat{\boldsymbol{\theta}}$ of the true parameter $\boldsymbol{\theta}$.
The complex CRB \cite{Vandenbos1994} states, 
$$
\text{cov}\big(g(\hat{\boldsymbol{\theta}})\big) \geq \frac{\partial g(\boldsymbol{\theta})}{\partial \boldsymbol{\theta}}\mathbf{I}(\boldsymbol{\theta})^{-1}\frac{\partial g(\boldsymbol{\theta})}{\partial \boldsymbol{\theta}}^H,
$$
with $
\mathbf{I}(\boldsymbol{\theta}) \triangleq \mathbb{E}\left[\frac{\partial\log\mathsf{L}}{\partial\boldsymbol{\theta}}\frac{\partial\log\mathsf{L}}{\partial\boldsymbol{\theta}}^H\right]
$ the Fisher Information Matrix (FIM), where $\mathsf{L}$ denotes the model likelihood, and $g$ is any complex differentiable vector function.
In particular, regarding the variance term of eq.~\eqref{eq:MSE_decomposed}, 
\begin{equation}
\Tr\Big(\text{cov}\big(\mathbf{h}(\hat{\boldsymbol{\theta}})\big)\Big) \geq \Tr\Big(\frac{\partial \mathbf{h}(\boldsymbol{\theta})}{\partial \boldsymbol{\theta}}\mathbf{I}(\boldsymbol{\theta})^{-1}\frac{\partial \mathbf{h}(\boldsymbol{\theta})}{\partial \boldsymbol{\theta}}^H\Big),
\label{eq:trace_inequality}
\end{equation}
with $
 \frac{\partial \mathbf{h}(\boldsymbol{\theta})}{\partial \boldsymbol{\theta}}
=
\big(\frac{\partial \mathbf{h}(\boldsymbol{\theta})}{\partial \theta_1},\dots,\frac{\partial \mathbf{h}(\boldsymbol{\theta})}{\partial \theta_{n_p}} \big).
$
A model independent expression for the FIM is provided in section~\ref{ssec:general_derivation}, and particularized in section~\ref{ssec:bound_sparse_channel} to the model of section:~\ref{ssec:model}. Finally, the bound is derived from eq.~\eqref{eq:trace_inequality} in section~\ref{ssec:general_bound}.

\subsection{General derivation}
\label{ssec:general_derivation}
First, notice that vectorizing eq.~\eqref{eq:observation_model}, the observation matrix $\mathbf{Y}$ follows a complex gaussian distribution,
$$
\text{vec}(\mathbf{Y}) \sim \mathcal{CN}\big(\underbrace{(\mathbf{X}^T\otimes \mathbf{W}^H) \mathbf{h}(\boldsymbol{\theta})}_{\boldsymbol{\mu}(\boldsymbol{\theta})},\underbrace{\sigma^2(\mathbf{Id}_{n_s}\otimes\mathbf{W}^H\mathbf{W})}_{\boldsymbol{\Sigma}} \big). $$
In that particular case, the Slepian-Bangs formula \cite{Slepian1954,Bangs1971} yields: 
\begin{equation}
\renewcommand*{\arraystretch}{1.8}
\arraycolsep=0.6pt
\begin{array}{rl}
\mathbf{I}(\boldsymbol{\theta}) &= 2\mathfrak{Re}\left\{\frac{\partial \boldsymbol{\mu}(\boldsymbol{\theta})}{\partial \boldsymbol{\theta}}^H\boldsymbol{\Sigma}^{-1}\frac{\partial \boldsymbol{\mu}(\boldsymbol{\theta})}{\partial \boldsymbol{\theta}}\right\}
\\
&= \frac{2\alpha^2}{\sigma^2}\mathfrak{Re}\left\{\frac{\partial \mathbf{h}(\boldsymbol{\theta})}{\partial \boldsymbol{\theta}}^H\mathbf{P}\frac{\partial \mathbf{h}(\boldsymbol{\theta})}{\partial \boldsymbol{\theta}}\right\}, 
\end{array}
\label{eq:FIM_gen}
\end{equation}
%with
%$
%\mathbf{S}(\boldsymbol{\theta}) \triangleq \frac{\partial \boldsymbol{\mu}(\boldsymbol{\theta})}{\partial \boldsymbol{\theta}}
%=
%\left(\frac{\partial \boldsymbol{\mu}(\boldsymbol{\theta})}{\partial \theta_1},\dots,\frac{\partial \boldsymbol{\mu}(\boldsymbol{\theta})}{\partial \theta_{n_p}} \right)
%$, often referred to as the sensibility matrix.
%Using elementary properties of the vectorization, we have
%$\boldsymbol{\mu}(\boldsymbol{\theta}) = (\mathbf{X}^T\otimes \mathbf{W}^H) \mathbf{h}(\boldsymbol{\theta})$, that implies
%$$
%\mathbf{S}(\boldsymbol{\theta})  = (\mathbf{X}^T\otimes \mathbf{W}^H)\frac{\partial \mathbf{h}(\boldsymbol{\theta})}{\partial \boldsymbol{\theta}}, 
%$$
with $\mathbf{P}\triangleq \frac{\sigma^2}{\alpha^2}(\mathbf{X}^*\otimes \mathbf{W})\boldsymbol{\Sigma}^{-1}(\mathbf{X}^T\otimes \mathbf{W}^H)$ where $\alpha^2 \triangleq \frac{1}{n_s} \text{Tr}(\mathbf{X}^H\mathbf{X})$ is the average transmit power per time step.
%which, injecting the expression of $\boldsymbol{\mu}(\boldsymbol{\theta}) $ becomes 
%$$
%\mathbf{I}(\boldsymbol{\theta}) = 2\mathfrak{Re}\left\{\frac{\partial \mathbf{h}(\boldsymbol{\theta})}{\partial \boldsymbol{\theta}}^H(\mathbf{X}^*\otimes \mathbf{W})\boldsymbol{\Sigma}^{-1}(\mathbf{X}^T\otimes \mathbf{W}^H)\frac{\partial \mathbf{h}(\boldsymbol{\theta})}{\partial \boldsymbol{\theta}}\right\}.
%$$
Note that the expression can be simplified to  $ \mathbf{P} = \frac{1}{\alpha^2}\big((\mathbf{X}^*\mathbf{X}^T)\otimes (\mathbf{W}(\mathbf{W}^H\mathbf{W})^{-1}\mathbf{W}^H)\big)$ using elementary properties of the Kronecker product. The matrix $\mathbf{W}(\mathbf{W}^H\mathbf{W})^{-1}\mathbf{W}^H$ is a projection matrix onto the range of $\mathbf{W}$. In order to ease further interpretation, assume that $\mathbf{X}^H\mathbf{X} = \alpha^2\mathbf{Id}_{n_s}$. This assumption means that the transmit power is constant during training time ($\left\Vert \mathbf{x}_i \right\Vert_2^2 = \alpha^2$, $\forall i$) and that pilots sent at different time instants are mutually orthogonal ($\mathbf{x}_i^H\mathbf{x}_j = 0$, $\forall i\neq j$). This way, $\frac{1}{\alpha^2}\mathbf{X}^*\mathbf{X}^T$ is a projection matrix onto the range of $\mathbf{X}^*$, and $\mathbf{P}$ can itself be interpreted as a projection, being the Kronecker product of two projection matrices \cite[p.112]{Steeb2011} (it is an orthogonal projection since $\mathbf{P}^H = \mathbf{P}$).

\subsection{Fisher information matrix for a sparse channel model}
\label{ssec:bound_sparse_channel}
Consider now the parametric channel model of section~\ref{ssec:model}, where $\mathbf{h} = \sum_{p=1}^P \mathbf{h}_p$, with $\mathbf{h}_p = c_p\mathbf{e}_t(\overrightarrow{u_{t,p}})^*\otimes \mathbf{e}_r(\overrightarrow{u_{r,p}})$. 

\noindent {\bf Intra-path couplings.} The derivatives of $\mathbf{h}$ with respect to parameters of the $p$th path $\boldsymbol{\theta}^{(p)}$ can be determined using matrix differentiation rules \cite{Petersen2008}:
\begin{itemize}[leftmargin=*,noitemsep,nolistsep]
\item Regarding the complex gain $c_p = \rho_p\mathrm{e}^{\mathrm{j}\phi_p}$, the model yields the expressions $\frac{\partial \mathbf{h}(\boldsymbol\theta)}{\partial \rho_p} = \frac{1}{\rho_p}\mathbf{h}_{p}$ and $\frac{\partial \mathbf{h}(\boldsymbol\theta)}{\partial \phi_p} = \mathrm{j}\mathbf{h}_{p}$.
\item Regarding the DoA, $\frac{\partial \mathbf{h}(\boldsymbol\theta)}{\partial \eta_{r,p}} =  \left(\mathbf{Id}_{n_t}\otimes \text{diag}(-\mathrm{j}\mathbf{A}_r^T\overrightarrow{v_{\eta_{r,p}}})\right)\mathbf{h}_{p}$ and $\frac{\partial \mathbf{h}(\boldsymbol\theta)}{\partial \psi_{r,p}} =  \left(\mathbf{Id}_{n_t}\otimes \text{diag}(-\mathrm{j}\mathbf{A}_r^T\overrightarrow{v_{\psi_{r,p}}})\right)\mathbf{h}_{p}$, where $\overrightarrow{v_{\eta_{r,p}}}$ and $\overrightarrow{v_{\psi_{r,p}}}$ are the unit vectors in the azimuth and elevation directions at $\overrightarrow{u_{r,p}}$, respectively.
\item Regarding the DoD, $\frac{\partial \mathbf{h}(\boldsymbol\theta)}{\partial \eta_{t,p}} =  \left(\text{diag}(\mathrm{j}\mathbf{A}_t^T\overrightarrow{v_{\eta_{t,p}}}) \otimes \mathbf{Id}_{n_r}\right)\mathbf{h}_{p}$ and $\frac{\partial \mathbf{h}(\boldsymbol\theta)}{\partial \psi_{t,p}} =  \left(\text{diag}(\mathrm{j}\mathbf{A}_t^T\overrightarrow{v_{\psi_{t,p}}}) \otimes \mathbf{Id}_{n_r}\right)\mathbf{h}_{p}$, where $\overrightarrow{v_{\eta_{t,p}}}$ and $\overrightarrow{v_{\psi_{t,p}}}$ are the unit vectors in the azimuth and elevation directions at $\overrightarrow{u_{t,p}}$, respectively.
\end{itemize}
Denoting $\frac{\partial\mathbf{h}}{\partial\boldsymbol{\theta}^{(p)}} \triangleq \left(\frac{\partial \mathbf{h}(\boldsymbol\theta)}{\partial \rho_p},\frac{\partial \mathbf{h}(\boldsymbol\theta)}{\partial \phi_p},\frac{\partial \mathbf{h}(\boldsymbol\theta)}{\partial \eta_{r,p}},\frac{\partial \mathbf{h}(\boldsymbol\theta)}{\partial \psi_{r,p}},\frac{\partial \mathbf{h}(\boldsymbol\theta)}{\partial \eta_{t,p}},\frac{\partial \mathbf{h}(\boldsymbol\theta)}{\partial \psi_{t,p}}\right)$, the part of the FIM corresponding to couplings between the parameters $\boldsymbol{\theta}^{(p)}$ (\emph{intra-path} couplings) is expressed as
\begin{equation}
\mathbf{I}^{(p,p)} \triangleq \frac{2\alpha^2}{\sigma^2}\mathfrak{Re}\left\{ \frac{\partial \mathbf{h}^H}{\partial \boldsymbol{\theta}^{(p)}}\mathbf{P}\frac{\partial \mathbf{h}}{\partial \boldsymbol{\theta}^{(p)}} \right\} .
\label{eq:FIM_sparse_channel}
\end{equation}

Let us now particularize this expression. First of all, in order to ease interpretations carried out in section~\ref{ssec:interpretation}, consider the case of \emph{optimal observation} conditions (when the range of $\mathbf{P}$ contains the range of $\frac{\partial\mathbf{h}(\boldsymbol{\theta})}{\partial{\boldsymbol{\theta}}}$). This allows indeed to interpret separately the role of the observation matrices and the antenna arrays geometries. Second, 
%let us focus on the diagonal blocks $\mathbf{I}^{(p,p)}$ of $\mathbf{I}(\boldsymbol{\theta})$, corresponding to couplings between parameters of the same path, or \emph{intra-path} couplings. 
consider for example the entry corresponding to the coupling between the departure azimuth angle $\eta_{t,p}$ and the arrival azimuth angle $\eta_{r,p}$ of the $p$th path. It is expressed under the optimal observation assumption as $\frac{2\alpha^2}{\sigma^2}\mathfrak{Re}\left\{\frac{\partial \mathbf{h}(\boldsymbol\theta)}{\partial \eta_{r,p}}^H\frac{\partial \mathbf{h}(\boldsymbol\theta)}{\partial \eta_{t,p}} \right\}$. Moreover, 
$$
\arraycolsep=0.6pt
\begin{array}{rl}
 \frac{\partial \mathbf{h}(\boldsymbol\theta)}{\partial \eta_{r,p}}^H\frac{\partial \mathbf{h}(\boldsymbol\theta)}{\partial \eta_{t,p}} 
=& \mathbf{h}_{p}^H  \Big( \text{diag}\big(\mathrm{j}\mathbf{A}_t^T\overrightarrow{v_{\eta_{t,p}}}\big)\otimes\text{diag}\big(\mathrm{j}\mathbf{A}_r^T\overrightarrow{v_{\eta_{r,p}}}\big) \Big) \mathbf{h}_{p}
\\
=&\frac{-\rho_p^2}{n_rn_t}\big(\mathbf{1}_{n_t}^T\mathbf{A}_t^T\overrightarrow{v_{\eta_{t,p}}}\big)\big(\mathbf{1}_{n_r}^T\mathbf{A}_r^T\overrightarrow{v_{\eta_{r,p}}}\big) =0,
\end{array}
$$
 since $\mathbf{A}_r\mathbf{1}_{n_r} = 0$ and $\mathbf{A}_t\mathbf{1}_{n_t} = 0$ by construction (because the antennas positions are taken with respect to the array centroid). This means that the parameters $\eta_{r,p}$ and $\eta_{t,p}$ are statistically uncoupled, i.e.\ \emph{orthogonal parameters} \cite{Cox1987}. Computing all couplings for $\boldsymbol{\theta}^{(p)}$ yields
%$$
%\begin{array}{rl}
% \mathbf{h}_{p}^H\mathbf{D}_{\eta_{r,p}}^H\mathbf{D}_{\eta_{r,p}}\mathbf{h}_{p}
%&=\frac{\rho_p^2}{n_r}\left\Vert\mathbf{A}_r^T\overrightarrow{v_{\eta_{r,p}}}\right\Vert_2^2,
%\end{array}
%$$
\begin{equation}
\renewcommand*{\arraystretch}{1.2}
%\resizebox{\textwidth}{!}{$
%~\hspace{-5cm}
\mathbf{I}^{(p,p)} = \frac{2\rho_p^2\alpha^2}{\sigma^2} 
\left(
\begin{array}{cccccc}
%\begin{smallmatrix}
\frac{1}{\rho_p^2}
&
0
&
\mathbf{0}_{1\times 2}
&
\mathbf{0}_{1\times 2}
\\ 
0
&
1
&
\mathbf{0}_{1\times 2}
&
\mathbf{0}_{1\times 2}
\\ 
\mathbf{0}_{2\times 1}
&
\mathbf{0}_{2\times 1}
&
\mathbf{B}_r
& 
\mathbf{0}_{2\times 2}
\\ 
\mathbf{0}_{2\times 1}
&
\mathbf{0}_{2\times 1}
&
\mathbf{0}_{2\times 2}
& 
\mathbf{B}_t
\\ 
\end{array}
%\end{smallmatrix}
\right),%$}
\label{eq:Ipp}
\end{equation}
where
\begin{equation}
\renewcommand*{\arraystretch}{1.7}
\mathbf{B}_x = 
\frac{1}{n_x}\left(\begin{array}{cc}
\left\Vert\mathbf{A}_x^T\overrightarrow{v_{\eta_{x,p}}}\right\Vert_2^2
& 
\overrightarrow{v_{\eta_{x,p}}}^T\mathbf{A}_x\mathbf{A}_x^T\overrightarrow{v_{\psi_{x,p}}}
\\ 
\overrightarrow{v_{\psi_{x,p}}}^T\mathbf{A}_x\mathbf{A}_x^T\overrightarrow{v_{\eta_{x,p}}}
& 
\left\Vert\mathbf{A}_x^T\overrightarrow{v_{\psi_{x,p}}}\right\Vert_2^2
\end{array}\right),
\label{eq:Bx}
\end{equation}
with $x \in \{r,t\}$. These expressions are thoroughly interpreted in section~\ref{ssec:interpretation}.

%Injecting the general expression of eq.~\eqref{eq:general_derivative} into eq.~\eqref{eq:FIM_gen}, the FIM entry at the $i$th row and $j$th column (which corresponds to the statistical coupling between the $i$th and $j$th parameters) takes the form
%\begin{equation}
%\mathbf{I}(\boldsymbol{\theta})_{ij} = \frac{2\alpha^2}{\sigma^2}\mathfrak{Re}\left\{ \mathbf{h}_{p(i)}^H\mathbf{D}_{\theta_i}^H\mathbf{P}\mathbf{D}_{\theta_j}\mathbf{h}_{p(j)} \right\}.
%\end{equation}
\noindent {\bf Global FIM.} Taking into account couplings between all paths, The global FIM is easily deduced from the previous calculations and block structured,
%\tiny
%$$
%\mathbf{I}(\boldsymbol{\theta}) =
%\renewcommand*{\arraystretch}{0.6} 
%\left(\begin{array}{cccc}
%\mathbf{I}^{(1,1)} & \mathbf{I}^{(1,2)}&\dots& \mathbf{I}^{(1,P)} \\
%\mathbf{I}^{(2,1)} & \mathbf{I}^{(2,2)}&& \\
%\svdots&&\sddots&\\
%\mathbf{I}^{(P,1)} & && \mathbf{I}^{(P,P)}
%\end{array}\right),
%$$
$$
\mathbf{I}(\boldsymbol{\theta}) =
\left(\begin{smallmatrix}
\mathbf{I}^{(1,1)} & \mathbf{I}^{(1,2)}&\dots& \mathbf{I}^{(1,P)} \\
\mathbf{I}^{(2,1)} & \mathbf{I}^{(2,2)}&& \\
\svdots&&\sddots&\\
\mathbf{I}^{(P,1)} & && \mathbf{I}^{(P,P)}
\end{smallmatrix}\right),
$$
\normalsize
where $\mathbf{I}^{(p,q)} \in \mathbb{R}^{6\times 6}$ contains the couplings between parameters of the $p$th and $q$th paths and is expressed $\mathbf{I}^{(p,q)} \triangleq \frac{2\alpha^2}{\sigma^2}\mathfrak{Re}\big\{ \frac{\partial \mathbf{h}}{\partial \boldsymbol{\theta}^{(p)}}^H\mathbf{P}\frac{\partial \mathbf{h}}{\partial \boldsymbol{\theta}^{(q)}} \big\} $.
The off-diagonal blocks $\mathbf{I}^{(p,q)}$ of $\mathbf{I}(\boldsymbol{\theta})$, corresponding to couplings between parameters of distinct paths, or \emph{inter-path} couplings, can be expressed explicitly (as in eq.~\eqref{eq:Ipp} for intra-path couplings). However, the obtained expressions are less prone to interesting interpretations, and inter-paths couplings have been observed to be negligible in most cases. They are thus not displayed in the present paper, for brevity reasons. Note that a similar FIM computation was recently carried out in the particular case of linear arrays \cite{Garcia2017}. However, the form of the FIM (in particular parameter orthogonality) was not exploited in \cite{Garcia2017}, as is done here in sections~\ref{ssec:interpretation} and ~\ref{ssec:experiments}. % More involved for inter-path couplings, but observed to be proportional to $|\mathbf{h}_{p(i)}^H\mathbf{h}_{p(j)}|$ the correlation between atomic channels.

\subsection{Bound on the variance}
\label{ssec:general_bound}

The variance of channel estimators remains to be bounded, using eq.~\eqref{eq:trace_inequality}. From eq.~\eqref{eq:FIM_gen}, the FIM can be expressed more conveniently only with real matrices as $\mathbf{I}(\boldsymbol{\theta}) = \frac{2\alpha^2}{\sigma^2}\bar{\mathbf{D}}^T\bar{\mathbf{P}}\bar{\mathbf{D}},$
with
$$
\bar{\mathbf{D}} \triangleq \left( \begin{array}{c}
\mathfrak{Re}\{\frac{\partial \mathbf{h}(\boldsymbol{\theta})}{\partial \boldsymbol{\theta}}\} \\
\mathfrak{Im}\{\frac{\partial \mathbf{h}(\boldsymbol{\theta})}{\partial \boldsymbol{\theta}}\}
\end{array} \right),\quad 
%\in \mathbb{R}^{2n_rn_t \times n_p},\quad
\bar{\mathbf{P}} \triangleq \left(\begin{array}{cc}
\mathfrak{Re}\{\mathbf{P}\} & -\mathfrak{Im}\{\mathbf{P}\} \\
\mathfrak{Im}\{\mathbf{P}\} & \mathfrak{Re}\{\mathbf{P}\}
\end{array}\right), %\in \mathbb{R}^{2n_rn_t \times 2n_rn_t},
$$
where $\bar{\mathbf{P}}$ is also a projection matrix.
%Notice that $\bar{\mathbf{P}}$ is also a projection matrix, since it can be verified straightforwardly that $\bar{\mathbf{P}}\bar{\mathbf{P}}=\bar{\mathbf{P}}$. 
Finally, injecting eq.~\eqref{eq:FIM_gen} into eq.~\eqref{eq:trace_inequality} assuming the FIM is invertible, gives for the relative variance
\begin{equation}
\renewcommand*{\arraystretch}{1.3}
\arraycolsep=0.6pt
\begin{array}{rl}
\Tr\big(\text{cov}\big(\mathbf{h}(\hat{\boldsymbol{\theta}})\big)\big).\Vert \mathbf{h}\Vert_2^{-2} &\geq \frac{\sigma^2}{2\alpha^2}\Tr\left(\bar{\mathbf{D}}(\bar{\mathbf{D}}^T\bar{\mathbf{P}}\bar{\mathbf{D}})^{-1}\bar{\mathbf{D}}^T\right).\Vert \mathbf{h}\Vert_2^{-2}\\
&\geq \frac{\sigma^2}{2\alpha^2}\Tr\left(\bar{\mathbf{D}}(\bar{\mathbf{D}}^T\bar{\mathbf{D}})^{-1}\bar{\mathbf{D}}^T\right).\Vert \mathbf{h}\Vert_2^{-2}\\
 &= \frac{\sigma^2}{2\alpha^2\Vert \mathbf{h}\Vert_2^{2}}n_p = \frac{3P}{\text{SNR}},
\end{array}
\label{eq:bound_variance}
\end{equation}
where the second inequality comes from the fact that $\bar{\mathbf{P}}$ being an orthogonal projection matrix, $\bar{\mathbf{P}} \leq \mathbf{Id} \Rightarrow \bar{\mathbf{D}}^T\bar{\mathbf{P}}\bar{\mathbf{D}} \leq \bar{\mathbf{D}}^T\bar{\mathbf{D}} \Rightarrow (\bar{\mathbf{D}}^T\bar{\mathbf{P}}\bar{\mathbf{D}})^{-1} \geq (\bar{\mathbf{D}}^T\bar{\mathbf{D}})^{-1} \Rightarrow \bar{\mathbf{D}}(\bar{\mathbf{D}}^T\bar{\mathbf{P}}\bar{\mathbf{D}})^{-1}\bar{\mathbf{D}}^T \geq \bar{\mathbf{D}}(\bar{\mathbf{D}}^T\bar{\mathbf{D}})^{-1}\bar{\mathbf{D}}^T$ (using elementary properties of the ordering of semidefinite positive matrices, in particular \cite[Theorem 4.3]{Baksalary1989}). The first equality comes from the fact that $\text{Tr}\big(\bar{\mathbf{D}}(\bar{\mathbf{D}}^T\bar{\mathbf{D}})^{-1}\bar{\mathbf{D}}^T\big) = \text{Tr}(\mathbf{Id}_{n_p}) = n_p$. Finally, the second equality is justified by $n_p = 6P$ considering the sparse channel model, and by taking $\text{SNR} \triangleq \tfrac{\alpha^2\Vert \mathbf{h}\Vert_2^{2}}{\sigma^2}$ (this is actually an optimal SNR, only attained with perfect precoding and combining). 

\noindent{\bf Optimal bound.} The first inequality in eq.~\eqref{eq:bound_variance} becomes an equality if an efficient estimator is used \cite{Kay1993}. Moreover, the second inequality is an equality if the condition $\text{im}\big(\frac{\partial\mathbf{h}(\boldsymbol{\theta})}{\partial\boldsymbol{\theta}}\big) \subset \text{im}\left( \mathbf{P} \right)$ is fulfilled (this corresponds to optimal observations, further discussed in section~\ref{ssec:interpretation}). Remarkably, under optimal observations, the lower bound on the relative variance is directly proportional to the considered number of paths $P$ and inversely proportional to the SNR, and does not depend on the specific model structure, since the influence of the derivative matrix $\bar{\mathbf{D}}$ cancels out in the derivation.
 
%  Third, it may guide the design of practical algorithms, as explained in the next section.

\noindent{\bf Sparse recovery CRB.} It is interesting to notice that the bound obtained here is similar to the CRB for sparse recovery \cite{Benhaim2010} (corresponding to an intrinsically discrete model), that is proportional to the sparsity of the estimated vector, analogous here to the number of paths.

%~\vspace{-5mm}
\section{Interpretations}
\label{ssec:interpretation}
The main results of sections~\ref{ssec:bound_sparse_channel} and \ref{ssec:general_bound} are interpreted in this section, ultimately guiding the design of efficient estimation algorithms.

\noindent{\bf Parameterization choice.} The particular expression of the FIM allows to assess precisely the chosen parameterization. First of all, $\mathbf{I}(\boldsymbol{\theta})$ has to be invertible and well-conditioned, for the model to be theoretically and practically identifiable \cite{Rothenberg1971,Kravaris2013}, respectively. As a counterexample, imagine two paths indexed by $p$ and $q$ share the same DoD and DoA, then proportional columns appear in $\frac{\partial\mathbf{h}(\boldsymbol{\theta})}{\partial{\boldsymbol{\theta}}}$, which implies non-invertibility of the FIM. However, it is possible to summarize the effect of these two paths with a single \emph{virtual} path of complex gain $c_p + c_q$ without any accuracy loss in channel description, yielding an invertible FIM. Similarly, two paths with very close DoD and DoA yield an ill-conditioned FIM (since the corresponding steering vectors are close to colinear), but can be merged into a single virtual path with a limited accuracy loss, improving the conditioning. Interestingly, in most channel models, paths are assumed to be grouped into clusters, in which all DoDs and DoAs are close to a principal direction \cite{Saleh1987, Wallace2002, Jensen2004}. Considering the MSE, merging close paths indeed decreases the variance term (lowering the total number of parameters), without increasing significantly the bias term (because their effects on the channel matrix are very correlated).
These considerations suggest dissociating the number of paths considered in the model $P$ from the number of \emph{physical} paths, denoted $P_{\phi}$, taking $P<P_{\phi}$ by merging paths. This is one motivation behind the famous virtual channel representation \cite{Sayeed2002}, where the resolution at which paths are merged is fixed and given by the number of antennas. The theoretical framework of this paper suggests to set $P$ (and thus the merging resolution) so as to minimize the MSE. 
%Assessing precisely the impact of the chosen $P$ on the bias/variance trade-off would thus be of great interest. 
A theoretical study of the bias term of the MSE (which should decrease when $P$ increases) could thus allow to calibrate models, choosing an optimal number of paths $P^*$ for estimation. Such a quest for $P^*$ is carried out empirically in section~\ref{ssec:experiments}.

\noindent{\bf Optimal observations.} The matrices $\mathbf{X}$ and $\mathbf{W}$ (pilot symbols and analog combiners) determine the quality of channel observation. Indeed, it was shown in section~\ref{ssec:general_bound} that the lowest CRB is obtained when 
$\text{im}\big(\frac{\partial\mathbf{h}(\boldsymbol{\theta})}{\partial\boldsymbol{\theta}}\big) \subset \text{im}\left( \mathbf{P} \right)$,
with $ \mathbf{P} = \frac{1}{\alpha^2}\big((\mathbf{X}^*\mathbf{X}^T)\otimes (\mathbf{W}(\mathbf{W}^H\mathbf{W})^{-1}\mathbf{W}^H)\big)$.
In case of sparse channel model, using the expressions for $\frac{\partial\mathbf{h}(\boldsymbol{\theta})}{\partial\boldsymbol{\theta}}$ derived above, this is equivalent to two distinct conditions for the training matrix:~ \vspace{-2mm} 
$$
\text{span}\left(\bigcup_{p=1}^P\Big\{\mathbf{e}_t(\overrightarrow{u_{t,p}}),\frac{\partial\mathbf{e}_t(\overrightarrow{u_{t,p}})}{\partial\eta_{t,p}},\frac{\partial\mathbf{e}_t(\overrightarrow{u_{t,p}})}{\partial\psi_{t,p}}\Big\}\right) \subset \text{im}(\mathbf{X}),
$$
~\vspace{-2mm}and for the analog combiners:~\vspace{-0mm}
$$
\text{span}\left(\bigcup_{p=1}^P\Big\{\mathbf{e}_r(\overrightarrow{u_{r,p}}),\frac{\partial\mathbf{e}_r(\overrightarrow{u_{r,p}})}{\partial\eta_{r,p}},\frac{\partial\mathbf{e}_r(\overrightarrow{u_{r,p}})}{\partial\psi_{r,p}}\Big\}\right) \subset \text{im}(\mathbf{W}),
$$
where
$\frac{\partial\mathbf{e}_x(\overrightarrow{u_{x,p}})}{\partial\xi_{x,p}} = \text{diag}(-\mathrm{j}\mathbf{A}_x^T\overrightarrow{v_{\xi_{x,p}}})\mathbf{e}_x(\overrightarrow{u_{x,p}})$ with $x \in \{r,t\}$ and $\xi \in \{\eta,\psi\}$. These conditions are fairly intuitive: to estimate accurately parameters corresponding to a given DoD (respectively DoA), the sent pilot sequence (respectively analog combiners) should span the corresponding steering vector and its derivatives (to ``sense'' small changes). To accurately estimate all the channel parameters, it should be met for each atomic channel.

\noindent{\bf Array geometry.} Under optimal observation conditions, performance limits on DoD/DoA estimation are given by eq.~\eqref{eq:Bx}. The lower the diagonal entries $\mathbf{B}_x^{-1}$, the better the bound. This implies the bound is better if the diagonal entries of $\mathbf{B}_x$ are large and the off-diagonal entries are small (in absolute value). Since the unit vectors $\overrightarrow{v_{\eta_{x,p}}}$ and $\overrightarrow{v_{\psi_{x,p}}}$ are by definition orthogonal, having $\mathbf{A}_x\mathbf{A}_x^T = \beta^2\mathbf{Id}$  with maximal $\beta^2$ is optimal, and yields uniform performance limits for any DoD/DoA. Moreover, in this situation, $\beta^2$ is proportional to $\frac{1}{n_x}\sum_{i=1}^{n_x}\left\Vert\overrightarrow{a_{x,i}}\right\Vert_2^2$, the mean squared norm of antenna positions with respect to the array centroid. Having a larger antenna array is thus beneficial (as expected), because the furthest antennas are from the array centroid, the larger $\beta^2$ is.

\noindent{\bf Orthogonality of DoA and DoD.} Section~\ref{ssec:bound_sparse_channel} shows that the matrix corresponding to intra-path couplings (eq.~\eqref{eq:Ipp}) is block diagonal, meaning that for a given path, parameters corresponding to gain, phase, DoD and DoA are mutually orthogonal. Maximum Likelihood (ML) estimators of orthogonal parameters are asymptotically independent \cite{Cox1987} (when the number of observations, or equivalently the SNR goes to infinity). %It means that parameters can be estimated, independently or jointly, with the same asymptotic performance, provided ML estimators are used in both cases. 
Classically, channel estimation in massive MIMO systems is done using greedy sparse recovery algorithms \cite{Mallat1993,Tropp2007,Bajwa2010}. Such algorithms can be cast into ML estimation with discretized directions, in which the DoD and DoA (coefficient support) are estimated \emph{jointly} first (which is costly), and then the gain and phase are deduced (coefficient value), iteratively for each path. Orthogonality between the DoD and DoA parameters is thus \emph{not} exploited by classical channel estimation methods. We propose here to exploit it via a sequential \emph{decoupled} DoD/DoA estimation, that can be inserted in any sparse recovery algorithm in place of the support estimation step, without loss of optimality in the ML sense. In the proposed method, one direction (DoD or DoA) is estimated first using an ML criterion considering the other direction as a nuisance parameter, and the other one is deduced using the joint ML criterion. Such a strategy is presented in algorithm~\ref{algo_summary}. It can be verified that lines $3$ and $6$ of the algorithm actually correspond to ML estimation of the DoA and joint ML estimation, respectively. The overall complexity of the sequential directions estimation is thus $\mathcal{O}(m+n)$, compared to $\mathcal{O}(mn)$ for the joint estimation with the same test directions. Note that a similar approach, in which DoAs for all paths are estimated at once first, was recently proposed \cite{Noureddine2017} (without theoretical justification). %The two strategies are compared in the next section.

\section{Preliminary experiment}
\label{ssec:experiments}
Let us compare the proposed sequential direction estimation to the classical joint estimation. This experiment must be seen as an example illustrating the potential of the approach, and not as an extensive experimental validation. 

\noindent {\bf Experimental settings.} Consider synthetic channels generated using the NYUSIM channel simulator \cite{Samimi2016} (setting $f=28$\,GHz, the distance between transmitter and receiver to $d=30$\,m) to obtain the DoDs, DoAs, gains and phases of each path. The channel matrix is then obtained from eq.~\eqref{eq:channel_model}, considering square Uniform Planar Arrays (UPAs) with half-wavelength separated antennas, with $n_t = 64$ and $n_r = 16$. Optimal observations are considered, taking both $\mathbf{W}$ and $\mathbf{X}$ as the identity. Moreover, the noise variance $\sigma^2$ is set so as to get an SNR of $10$\,dB. Finally, the two aforementioned direction estimation strategies are inserted in the Matching Pursuit (MP) algorithm \cite{Mallat1993}, discretizing the directions taking $m=n=2,500$, and varying the total number $P$ of estimated paths.
\begin{algorithm}[tb]
\caption{Sequential direction estimation (DoA first)}
\begin{algorithmic}[1] 
\STATE Choose $m$ DoAs to test: $\{\overrightarrow{u_{1}},\dots,\overrightarrow{u_{m}}\}$
\STATE Build the matrix $\mathbf{K}_r = \Big(\frac{\mathbf{W}^H\mathbf{e}_r(\overrightarrow{u_{1}})}{\left\Vert \mathbf{W}^H\mathbf{e}_r(\overrightarrow{u_{1}})\right\Vert_2}|\dots|\frac{\mathbf{W}^H\mathbf{e}_r(\overrightarrow{u_{n}})}{\left\Vert \mathbf{W}^H\mathbf{e}_t(\overrightarrow{u_{n}})\right\Vert_2}\Big) $ %and normalize its columns
\STATE Find the index $\hat{i}$ of the maximal entry of $\text{diag}(\mathbf{K}_r^H\mathbf{Y}\mathbf{Y}^H\mathbf{K}_r)$,\newline set $\hat{\overrightarrow{u_t}} \leftarrow \overrightarrow{u_{\hat{i}}}$ (\textcolor{black}{$\mathcal{O}(m)$ complexity})
\STATE Choose $n$ DoDs to test: $\{\overrightarrow{v_{1}},\dots,\overrightarrow{v_{n}}\}$
\STATE Build the matrix $\mathbf{K}_t = \Big(\frac{\mathbf{X}^H\mathbf{e}_t(\overrightarrow{v_{1}})}{\left\Vert \mathbf{X}^H\mathbf{e}_t(\overrightarrow{v_{1}})\right\Vert_2}|\dots|\frac{\mathbf{X}^H\mathbf{e}_t(\overrightarrow{v_{n}})}{\left\Vert \mathbf{X}^H\mathbf{e}_t(\overrightarrow{v_{n}})\right\Vert_2}\Big) $ %and normalize its columns
\STATE Find the index $\hat{j}$ of the maximal entry of $\mathbf{e}_r(\overrightarrow{u_{\hat{i}}})^H\mathbf{Y}\mathbf{K}_t$,\newline set $\hat{\overrightarrow{u_t}} \leftarrow \overrightarrow{v_{\hat{j}}}$ (\textcolor{black}{$\mathcal{O}(n)$ complexity})
\end{algorithmic}
\label{algo_summary}
\end{algorithm}

\begin{table}[b]                                         
\def\arraystretch{1.0}                                           
\begin{tabularx}{\columnwidth}{XXXXX} 
\toprule                                                                              
&\multicolumn{2}{l}{~\hspace{1.0mm}Joint estimation}& \multicolumn{2}{l}{~\hspace{-2.5mm}Sequential estimation} \\
 & rMSE  & Time & rMSE & Time \\
\midrule                                                                                                           
\multicolumn{1}{r}{$P=5$}& 0.077 & 1.24 & 0.092 & 0.11 \\                                                 
\multicolumn{1}{r}{$P=10$}&0.031 & 2.40 & 0.039 & 0.16 \\                                                 
%\multicolumn{1}{r}{$P=15$}&0.14 & 3.56 & 0.16& 0.20 \\                                                 
\multicolumn{1}{r}{$P=20$}&\bf 0.017& 4.66 & \bf 0.021 & 0.24 \\
\multicolumn{1}{r}{$P=40$}&0.025& 9.50 & 0.023 & 0.42 \\
\bottomrule                                                                                                      
\end{tabularx}   
~\vspace{-3mm}                                        
\caption{Relative MSE and estimation time (in seconds), in average over $100$ channel realizations, the lowest rMSE being shown in bold.}                                
\label{tab:results}                              
\end{table}

\noindent {\bf Results.} Table~\ref{tab:results} shows the obtained relative MSE and estimation times (Python implementation on a laptop with an Intel(R) Core(TM) i7-3740QM CPU @ 2.70 GHz). First of all, for $P=5,10,20$, the estimation error decreases and the estimation time increases with $P$, exhibiting a trade-off between accuracy and time. However, increasing $P$ beyond a certain point seems useless, since the error re-increases, as shown by the MSE for $P=40$, echoing the trade-off evoked in section~\ref{ssec:general_bound}, and indicating that $P^*$ is certainly between $20$ and $40$ for both methods in this setting.
Finally, for any value of $P$, while the relative errors of the sequential and joint estimation methods are very similar, the estimation time is much lower (between ten and twenty times) for sequential estimation. This observation validates experimentally the theoretical claims made in the previous section. 

%\section{Experiments}
%~\vspace{-4.3mm}
\section{Conclusions and perspectives}
In this paper, the performance limits of massive MIMO channel estimation were studied. To this end, training based estimation with a physical channel model and an hybrid architecture was considered. The Fisher Information Matrix and the Cram\'er-Rao bound were derived, yielding several results. The CRB ended up being proportional to the number of parameters in the model and independent from the precise model structure. The FIM allowed to draw several conclusions regarding the observation matrices and the arrays geometries. Moreover, it suggested computationally efficient algorithm which are asymptotically as accurate as classical ones.

This paper is obviously only a first step toward a deep theoretical understanding of massive MIMO channel estimation. Apart from more extensive experimental evaluations and optimized algorithms, a theoretical study of the bias term of the MSE would be needed to calibrate models, and the interpretations of section~\ref{ssec:interpretation} could be leveraged to guide system design.

%\begin{itemize}[leftmargin=*,noitemsep,nolistsep] 
%\item 
%First of all, algorithmic consequences of this work should be evidenced. Indeed, algorithms based on independent ML estimation of the DoD and DoA, as suggested in section~\ref{ssec:interpretation}, remain to be precisely stated and empirically assessed. This would allow to quantify the asymptotic statement we made in this paper. Modification of matching pursuit for the support estimation step. In the future, we could adapt to more evolved sparse recovery algorithms. We can also imagine departing from the sparse recovery world, for example with iterative algorithms instead of ones based on discretizing the direction spaces.
%\item 
%Second,

%\item Finally, the framework introduced in this paper could also be used to assess models with different structures. For example, we made in this paper the planar waves assumption. However, for very large arrays or very close transmitter and receiver, the assumption could be inaccurate. We can thus wonder to what extent it would be beneficial to consider spherical waves \cite{Jiang2005,Zhou2015}. Indeed, this would necessarily lower the bias, but how much supplementary variance would it introduce?
%\end{itemize}

\noindent {\bf Acknowledgments.} The authors wish to thank Matthieu Crussi\`ere for the fruitful discussions that greatly helped improving this work.

%\vfill\pagebreak

\bibliographystyle{IEEEbib}
\bibliography{biblio_mimo}

\end{document}